\documentstyle[preprint,aps,amsmath,epsfig]{revtex}

\newcommand{\amin}{\alpha_{\rm min}}
\newcommand{\amax}{\alpha_{\rm max}}
\newcommand{\den}{{\rm den}\:}
\newcommand{\omegacon}{\widetilde{\omega}}

\begin{document}

\tighten

\preprint{} 
 
\title{Hofstadter Rules and Generalized Dimensions of the Spectrum of  Harper's Equation}

\author{Andreas R\"udinger and Fr\'ed\'eric Pi\'echon} 

\address{Institut f\"ur Theoretische und Angewandte Physik\\ 
Universit\"at Stuttgart \\
Pfaffenwald 57 \\ 
70550 Stuttgart, Germany} 

\maketitle

\begin{abstract}
We consider the Harper model which describes two dimensional Bloch electrons 
in a magnetic field. 
For irrational flux through the unit-cell the corresponding energy spectrum is
known to be a Cantor set with multifractal properties. 
In order to relate the maximal and
minimal fractal dimension of the spectrum of Harper's equation to the
irrational number involved, we combine 
a refined version of the Hofstadter rules with results from
semiclassical analysis and tunneling in phase space.  
For quadratic irrationals $\omega$ with  continued fraction expansion 
$\omega = [0;\overline{n}]$ the maximal fractal dimension exhibits oscillatory
behavior as a function of $n$, which can be explained by the structure of the
renormalization flow. The asymptotic behavior of the minimal fractal dimension
is given by $\amin \sim {\rm const.} \ln n / n$. 
As the generalized dimensions can be related to the anomalous diffusion 
exponents of an initially  localized wavepacket, our results imply that the 
time evolution of high order moments $\langle r^{q} \rangle, \: q \to
\infty$ is sensible to the parity of $n$. 
\end{abstract}

\vskip 3cm  
{PACS numbers: 05.90.+m, 61.44.Fw, 71.90.+q}

Submitted to {\it Journal of Physics A} (version of \today)

\vskip3cm 

Originally conceived to describe Bloch electrons in a magnetic field 
\cite{Harper}, Harper's equation 
\begin{equation}
\phi_{n+1} +\phi_{n-1} + 2 \cos(2\pi \omega n +\nu) \phi_{n} = E \phi_{n}    
\label{harper}
\end{equation}
has become a subject of its own right. 
The parameter $\omega$ is the ratio between the magnetic flux per unit cell
and the flux quantum, $\phi_{n}$ is related to the wavefunction of the 
Bloch electron and $\nu$ is a phase. 
Harper-like models are of great interest in a wide variety of physical 
contexts (integer quantum Hall effect \cite{Thouless0}, 
superconducting networks \cite{Pannetier}, electrons 
in superlattices \cite{Gerhardts}) and have given rise to strong interplay between physics 
and mathematics, as e.g. in the fields of semiclassics \cite{Rammal} , non
commutative geometry \cite{Bellissard0}, and, quite recently, quantum groups 
\cite{Wiegmann}. 
Experimental investigations on superconducting networks permit a direct
observation of the ground state energy as a function of the magnetic flux \cite{Pannetier}. 
Furthermore, the resolution of the fine structure of Hofstadter's famous
butterfly have been made possible by measuring the magnetoresistance
oscillations in superlattices \cite{Gerhardts,Schlosser}.  

During the last years there has been great interest in the generalized
(R\'enyi) dimensions \cite{Halsey} of the spectral measure of this 
equation and similar quasiperiodic tight-binding hamiltonians 
\cite{Kohmoto1,Tang,Kohmoto2,Zheng,Hiramoto,Ikezawa,Fred1,Rudinger}.  
For the Harper equation it was conjectured that the fractal dimension $D_{q=0}$
is exactly 1/2 for typical irrational numbers \cite{Ikezawa,Bell}. 
Due to the Thouless property \cite{Thouless,Last,Tan}, however, there is a simple argument providing
strong evidence for $D_{0}$ being strictly smaller than $1/2$ \cite{remark}. 
Recently a statistical theory has been presented to explain the numerically 
observed behavior 
of $D_{0}$ as a function of the irrational number $\omega = [0;\overline{n}]$
\cite{Wilkinson2}. 
Until now there is no detailed investigation how the generalized dimensions
$D_{q}$ for a generic $q$ depend on the incommensurability $\omega$. 
However, the generalized dimensions 
play an important role for explaining the anomalous
 diffusion properties of wavepackets found in incommensurable models. 
Quite recently, the link between $D_{q}$ and the anomalous diffusion 
$\langle r^{q} \rangle  \sim t^{q\sigma_{q}}$ has been made explicit 
by the relation $\sigma_{q}=
D_{1-q}$, which has been verified numerically \cite{Fred2}.

In this paper we deal with the problem of the maximal and minimal fractal
dimensions $\amin = D_{+\infty}$ and $\amax = D_{-\infty}$. This is intended
to be a  first step to general values of $q$. 
While $\amin$ (as $D_{0}$) exhibits a monotonic behavior as a function of $n$,
we find that $\amax$ shows strongly pronounced even-odd oscillations that can
be explained by a refinement of the Hofstadter rules \cite{Hofstadter}.   
Similar behavior of $\amin$ and $\amax$ has been found for a tight-binding
Hamiltonian associated with substitution sequences using the trace map
approach \cite{Holzer1,Holzer2}. 
 
The paper is organized as follows: We begin by reviewing the Hofstadter rules
(1), then we consider the multifractal properties of Harper's equation for the
case of the golden and silver mean (2) and finally pass to more general
quadratic irrational numbers (3), where the minimal and maximal 
fractal dimensions of the related spectrum will be discussed in detail. 

\section{Hofstadter rules}

The starting point of our considerations is the hierarchical clustering of the
spectrum of  Harper's equation (\ref{harper}).  
By inspecting the band structure  of this equation  as a function of $\omega$, 
Hofstadter
\cite{Hofstadter} noticed that for a given $\omega$ it consists of three parts, one central cluster
(S) and two side clusters (R) that are closely related to the 
full spectra of Harper's equation with  renormalised values of $\omega$,
namely $R(\omega)$ and $S(\omega)$ with 
\begin{equation}
  R(\omega)= \begin{cases}
\{\frac{1}{\omega}\} & \text{if  $\quad 0<\omega < \frac{1}{2}$},\\
\{\frac{1}{1-\omega}\} & \text{if  $\quad \frac{1}{2} < \omega <1 $}
\end{cases}
\end{equation}
and 
\begin{equation}
  S(\omega)= \begin{cases}
\{\frac{\omega}{1-2\omega}\} & \text{if  $\quad 0<\omega < \frac{1}{2}$},\\
\{\frac{1-\omega}{2\omega-1}\} & \text{if  $\quad  \frac{1}{2} < \omega <1 $},
\end{cases}
\end{equation}
where $\{\cdot\}$ denotes the fractional part (for an illustration see
\cite{Hofstadter,Bell}). 

These clustering rules, denoted as Hostadter rules, have been derived
afterwards by different methods \cite{Bell,Wilkinson1}. 

Until now, however, for practical calculations, it has always been assumed
that the side clusters and the central cluster are simply rescaled versions 
of the full spectrum for the renormalized $\omega$. 
Furthermore only the case of the
golden and silver mean ($\omega_{\rm gold} = [0;\overline{1}]$, $\omega_{\rm
  silver} = [0;\overline{2}]$) has been considered. These particular values of
$\omega$ are fixed points of both maps, $R$ and $S$,  and therefore
self-consistent equations for the fractal dimension $D_{0}$ can be obtained
easily \cite{Bell}.  

In order to refine the assumption that the side and central clusters are
merely homogeneously rescaled versions of the full spectrum for the
renormalized $\omega$, we rewrite Hofstadter's clustering rules in a more
quantitative way. Consider the spectrum of a rational approximant $\omega =
p/q$ with $q$ bands. The partition of these bands on the lower side cluster,
the central cluster and the upper side cluster is given according to 
$\den \omega = q = \den R(\omega) + \den S(\omega) + \den R(\omega)$, where
"$\den $" denotes the denominator. 
Therefore we can write the $q$ energy values $E_{i}$ (upper or lower band edge,
for large $q$, in which we are interested,  the bandwidths go to zero, so
there is no difference) in the following form \cite{thesepiechon}: 
\begin{equation}
  E_{i}(\omega) = \begin{cases} f_{\omega}^{-}(E_{i}(R(\omega))) & \text{ for  }  i =   1,\cdots, \den R(\omega) \\
f_{\omega}^{0}(E_{i - \den R(\omega)}(S(\omega))) & \text{ for  } i = \den
R(\omega) + 1, \cdots, \den R(\omega) + \den S(\omega) \\
f_{\omega}^{+}(E_{i-\den R(\omega) - \den S(\omega)}(R(\omega))) & \text{ for
  }i =  \den R(\omega) + \den S(\omega) + 1,\cdots, q.
\end{cases}
\label{hofstadter}
\end{equation}
The functions $f_{\omega}^{\pm,0}(E)$ are the fingerprint of Hof\-stadter's
rules and are essential for calculating the fractal dimensions of the spectrum
for a given $\omega$. 
The irrational number $\omega$ can be developped in a continued fraction. Its
truncation at the $k$th level yields the approximant of $\omega$ of 
generation $k$.  
An essential point is that $f_{\omega}^{\pm,0}(E)$ only depend weakly  on the
generation of the approximant of $\omega$. 

For a quasiperiodic tight-binding model where the Hofstadter rules also
apply, the functions $f^{\pm,0}(E)$ have been calculated in a linear
approximation by means of a  
renormalization approach for the case of the golden mean. Since $\omega_{\rm
  gold}$ is invariant under $S$ and $R$, the Hof\-stadter rules close up after
the first step, leading to a self-consistent equation for the fractal
dimensions \cite{Fred1}. 

A generic quadratic irrational number is not invariant under $R$ and $S$, but successive applications of $R$ and/or $S$ 
involve a finite number of different irrational numbers. This can be seen, if
the renormalization equations for $\omega$ are rewritten for the continued
fraction expansion $\omega = [0;a_{1},a_{2},\cdots]$: 
\begin{equation}
R(\omega) = \begin{cases}
[0; a_{2},a_{3},a_{4}, \cdots] & \text{if $ a_{1} > 1$} , \\ {}
[0;  a_{3},a_{4},a_{5}, \cdots] & \text{if $ a_{1} = 1$}
\end{cases}
\label{equ:R-rules}
\end{equation}
and 
\begin{equation}
  S(\omega)=\begin{cases}
[0;a_{1}-2,a_{2},a_{3},\cdots] & \text{if $a_{1}>2 $}, \\  {}
[0;a_{3},a_{4},a_{5},\cdots] & \text{if $ a_{1} = 2 $},\\  {}
[0;a_{2}-1,a_{3},a_{4},\cdots] & \text{if $ a_{1}=1 , a_{2}>1 $}, \\ {}  
[0;a_{4},a_{5},a_{6},\cdots] & \text{if $a_{1}=1, a_{2}=1$}.
\end{cases}
\label{equ:S-rules}
\end{equation}

Since quadratic numbers have periodic continued fraction expansions, only a 
finite number of different irrationals will occur by applying iteratively
$R$ and $S$ to $\omega$.  
Thus, once  the functions $f_{\omega}(E)$ are known for all $\omega$ 
involved in the renormalization flow of the initial $\omega$, 
the multifractal properties of the spectrum are determined.  

We will restrict ourselves to quadratic numbers of the form $\omega = [0;
\overline{n}]$. Examples of  the renormalization flow are given in 
the following schematic representation for $n=3,4$: 

\unitlength0.7cm 
\begin{picture}(16,10.5)
\put(1,0.5){
\unitlength0.7cm 
\begin{picture}(10,10)
\put(4.2,8.8){$[0;\overline{3}]$} 
\put(4,5.8){$[0;1,\overline{3}]$}
\put(4,2.8){$[0;2,\overline{3}]$}
\put(4.5,8.5){\vector(0,-1){2}}
\put(4.8,8.5){\line(0,-1){2}}
\put(4.5,5.5){\vector(0,-1){2}}
\put(4.8,5.5){\line(0,-1){2}}
\put(4.5,2.5){\line(0,-1){2}}
\put(4.8,2.5){\line(0,-1){2}}
\put(4.8,6.5){\line(1,0){1.5}}
\put(4.8,3.5){\line(1,0){1.7}}
\put(4.8,0.5){\line(1,0){1.9}}
\put(6.3,6.5){\line(0,1){2.3}}
\put(6.5,3.5){\line(0,1){5.5}}
\put(6.7,0.5){\line(0,1){8.7}}
\put(4.5,0.5){\line(-1,0){2}}
\put(2.5,0.5){\line(0,1){8.5}}
\put(6.3,8.8){\vector(-1,0){1.0}}
\put(6.5,9.0){\vector(-1,0){1.2}}
\put(6.7,9.2){\vector(-1,0){1.4}}
\put(2.5,9){\vector(1,0){1.5}}
\put(3.6,7.5){S}
\put(5.2,7.5){R}
\put(3.6,4.5){S}
\put(5.2,4.5){R}
\put(3.6,1.5){S}
\put(5.2,1.5){R}
\end{picture}}
\put(11,0.5){
\unitlength0.7cm 
\begin{picture}(10,10)
\put(4.2,8.8){$[0;\overline{4}]$} 
\put(4,5.8){$[0;2,\overline{4}]$}
\put(4.5,8.5){\vector(0,-1){2}}
\put(4.8,8.5){\line(0,-1){2}}
\put(4.5,5.5){\line(0,-1){2}}
\put(4.8,5.5){\line(0,-1){2}}
\put(4.8,6.5){\line(1,0){1.5}}
\put(4.8,3.5){\line(1,0){1.7}}
\put(6.3,6.5){\line(0,1){2.3}}
\put(6.5,3.5){\line(0,1){5.5}}
\put(4.5,3.5){\line(-1,0){1.7}}
\put(2.8,3.5){\line(0,1){5.5}}
\put(6.3,8.8){\vector(-1,0){1.0}}
\put(6.5,9.0){\vector(-1,0){1.2}}
\put(2.8,9){\vector(1,0){1.3}}
\put(3.6,7.5){S}
\put(5.2,7.5){R}
\put(3.6,4.5){S}
\put(5.2,4.5){R}
\end{picture}}
\end{picture}

The structure of these diagrams depends on the parity of $n$: 
For $n$ odd, there are $n$ irrational numbers involved and $n+1$ cycles occur,
$n$ being of the form $S^{j}R$ and one of the form $S^{n}$. 
For $n$ even, there are $n/2$ irrationals and $n/2$ cycles occur,  
$n/2-1$ of the form $S^{j}R$ and one of the form $S^{n/2}$. \\

Each of these cycles do only leave invariant the exact irrational number
$\omega$. Applied to an approximant of $\omega$ of generation $k$, an
approximant of $\omega$, but of lower generation $k'<k$ is obtained, as is
immediately obvious by inspecting (\ref{equ:R-rules}) and
(\ref{equ:S-rules}). Thus we can assign to each cycle  a "generation loss"
$d:=k-k'$ (cf. Fig. \ref{tableau}).  

\section{Golden and silver number}

\begin{figure}
\centerline{\epsfig{figure=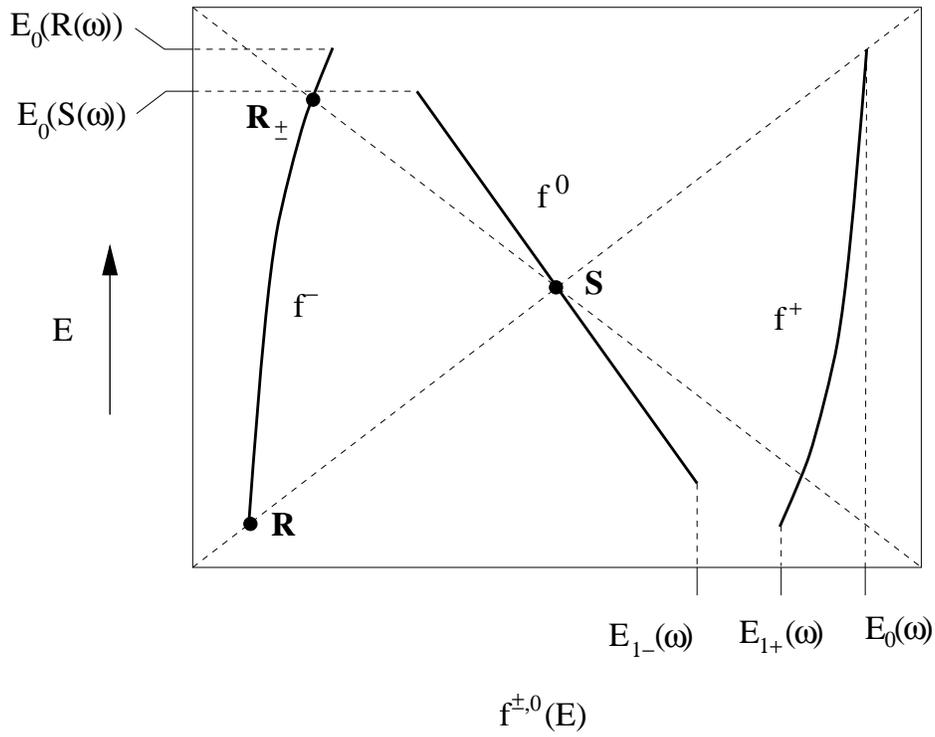,width=10cm,angle=270}}
\caption{Schematic representation of $f^{\pm,0}(E)$. The points ${\bf R}$ and
  ${\bf S}$ are the fixed points of $f^{\pm,0}(E)$. The point ${\bf R}_{\pm}$
corresponds to the two-orbit $E \to -E \to E$.}
\label{fig:fE}
\end{figure}

For the sake of simplicity, we will begin by dealing with the problem for the case $\omega = \omega_{\rm
  gold, silver}$, which are fixed points of the Hofstadter rules. 
The mapping $f^{\pm,0}_{\omega}(E)$ has two different fixed points, denoted as
$\bf R$ and $\bf S$ in Fig. \ref{fig:fE}, which correspond to the edges
respectively to the center of the spectrum.  
Assuming in a first approximation that the scaling properties of the
spectrum are  well described by the contraction  
factors at the centers and at the edges, 
$z_{S} = - \frac{{\rm d}f^{0}}{{\rm d}E}|_{\rm center}$ and 
$z_{R} =\frac{{\rm d}f^{\pm}}{{\rm d}E}|_{\rm edges}$, we find that after $k$ 
steps of renormalization, the bandwidhts scale with $\Delta \sim z^{k}$ while
the system size scales with $N \sim \omegacon^{-k\cdot d}$, where $\omegacon$
is the algebraically conjugate to $\omega$ ($\omegacon =
\frac{1}{2}(n+\sqrt{n^{2}+4})$ for $\omega = [0;\overline{n}] = \frac{1}{2}(-n
+ \sqrt{n^{2}+4})$).  

With $\Delta \sim N^{-1/\alpha}$ we obtain the two local scaling 
exponents at the edges and the center of the band:  
\begin{equation}
\alpha_{S} = - \frac{d(S) \ln \omegacon}{\ln z_{S}}   
\end{equation}
and
\begin{equation}
\alpha_{R} = - \frac{d(R) \ln \omegacon}{\ln z_{R}},     
\end{equation}
where $d(S)$ and $d(R)$ are the generation losses of the cycles $R$ and $S$.

Furthermore, within this approximation a self-consistent equation for
$\tau(q) = (q-1)D_{q}$, the Legendre transformed of $f(\alpha)$,  
can be given directly: 
\begin{equation}
  \frac{2\omega^{d(R)q}}{z_{R}^{\tau}} + \frac{\omega^{d(S)q}}{z_{S}^{\tau}} =
  1, \qquad \omega = \omega_{\rm gold}, \omega_{\rm silver}.  
\label{equ:2z}
\end{equation}
For the case of the golden number, this equation has recently been used to
derive the multifractal properties of a 
tight-binding Hamiltonian associated with the Fibonacci sequence in 
the limit of strong modulation \cite{Fred1}. Despite of being a good first
approximation, the corresponding $f(\alpha)$ has two serious drawbacks: First,
$f(\amin)$ or $f(\amax)$ (depending on the parameter) does not vanish in
contradiction to numerical simulations and indications from the trace map
approach \cite{Kohmoto2}.  
This problem is due to the
oversimplification of $f^{\pm,0}$ by three linear pieces, yielding a
artificial degeneration of the widths of the bands ($2^{n}$ bands of
$\omegacon^{n}$ ones have the same widths), which leads to $f(\alpha_{R}) \neq
0$. 

Second, in this approach the extremal scaling properties are
found at the band edges or at the center of the spectrum. This is,  however,  
by far  
not the general case. It is true for $\omega_{\rm gold}$ for both the Harper
model and the  quasiperiodic Hamiltonian mentioned above, 
but neither for $\omega_{\rm
  silver}$, nor for higher irrational numbers $\omega = [0;\overline{n}], n \ge
3$.   

In order to construct an approach eliminating these two problems, we consider
the mapping of the spectrum by the renormalization flow as a discrete dynamical
system. For $\omega_{\rm gold}$ and $\omega_{\rm silver}$ the spectral 
measure is given by the invariant measure of the map 
$f^{\pm,0}_{\omega}(E)$. 
The band edges and the band center are attracting fixed points of 
$f^{\pm,0}_{\omega}(E)$, thus giving rise to nontrivial scaling behavior as we
have seen above. 
To get accurate approximations of the $f(\alpha)$-curves, however, we have to
take into account higher periodic orbits of the underlying dynamical systems. 
The simplest approximation is to include the orbit $R_{+}R_{-}$ of period
two. The energies of the two (symmetric) fixed points are given by  
$f^{\pm}(E) = -E$.   
Therefore we find for the corresonding scaling factor 
\begin{equation}
  \alpha_{R_{+}R_{-}} = - \frac{d(R_{+}R_{-})\ln \omegacon}{\ln z_{R_{+}} + \ln
    z_{R_{-}}} = - \frac{ d(R) \ln \omegacon}{\ln z_{R_{\pm}}},  
\end{equation}
where $z_{R_{\pm}}$ is the local contraction factor at the fixed points of
order two:  
\begin{equation}
  z_{R_{\pm}} = \left. \frac{{\rm d} f^{\pm,0}(E)}{{\rm d} E}\right|_{E =
    -f^{\pm}(E)}. 
\end{equation}
Taking into account the three contraction factors $z_{R}, z_{S}, z_{R_{\pm}}$ 
(denoted as $3z$-model in the following) we arrive at the following 
self-consistent equation for $\tau(q)$: 
\begin{equation}
  \frac{\omega^{d(R)q}}{z_{R}^{\tau}} +
  \frac{\omega^{d(R)q}}{z_{R_{\pm}}^{\tau}} + 
\frac{\omega^{d(S)q}}{z_{S}^{\tau}} =
  1, \qquad \omega = \omega_{\rm gold}, \omega_{\rm silver}.  
\label{equ:3z}
\end{equation}
If $z_{R_{\pm}}$ equals  $z_{R}$  equation (\ref{equ:3z}) reduces to
(\ref{equ:2z}).  

The minimal and maximal dimensions for the $3z$-model are given by 
$\alpha_{\rm min (max)} =\min(\max)\{\alpha_{R},  \alpha_{S}, \alpha_{R_{\pm}}
\}$. By performing the Legendre transformation of (\ref{equ:3z}) we find the
correct behavior $f(\alpha_{\rm min, max}) = 0$. 
While for the golden number $\amax = \alpha_{S} > \alpha_{R_{\pm}} >
\alpha_{R}= \amin$, i.e. the maximal scaling exponent occurs at the center of
the spectrum, for the silver number the maximal dimension is given by $\amax =
\alpha_{R_{\pm}} > \alpha_{S}> \alpha_{R} = \amin$, i.e. the bands of maximal widths are not at
the center of the spectrum, but at the energy given by $f^{\pm}(E) = -E$. 
Therefore a linear approximation of $f_{\omega}^{\pm,0}(E)$
(i.e. two scaling factors, one for the central, and one for the side bands)
is not sufficent for explaining $\amin$ and $\amax$, even in the case of 
$\omega_{\rm  silver}$. 
  
\begin{figure}
\centerline{\epsfig{figure=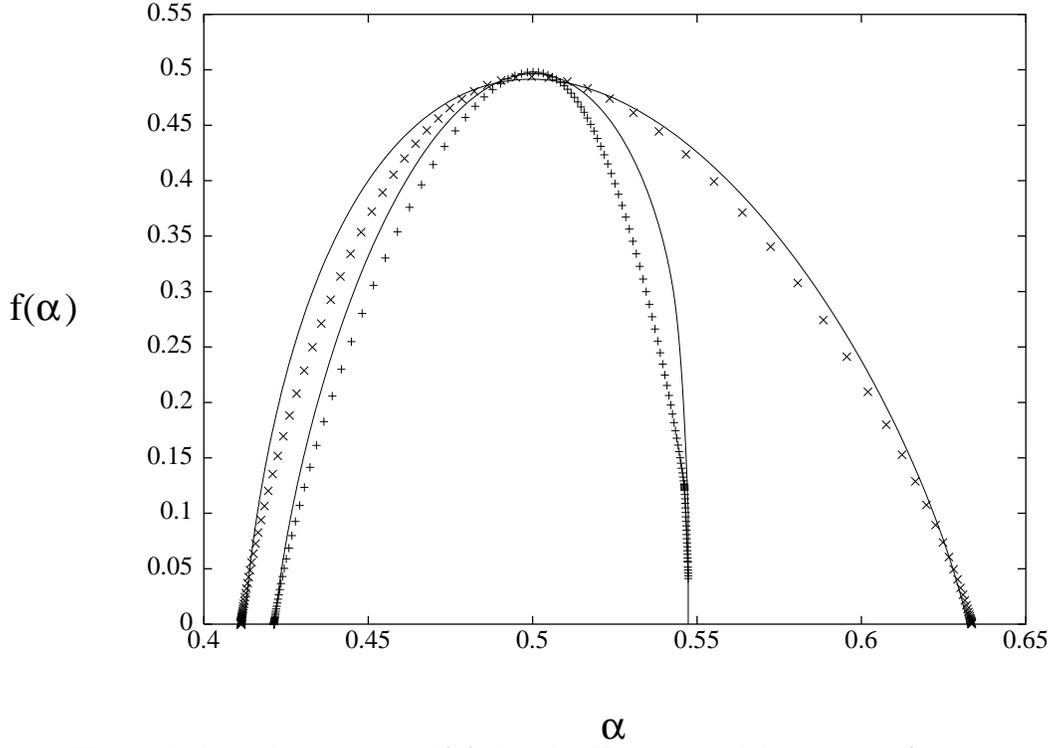,width=10cm,angle=270}}
\caption{The multifractal spectrum $f(\alpha)$ for the Harper model for
  $\omega_{\rm gold}$ (narrow curve) and $\omega_{\rm silver}$ (large curve) 
compared to the  $3z$-model. The $f(\alpha)$-curves obtained from diagonalization and the
  thermodynamic formalism are drawn with symbols, the $f(\alpha)$ obtained
  from Legendre-Transformation of (\ref{equ:3z}) are depicted with 
  lines ($z_{R}^{{\rm gold}} = 0.103$, $z^{\rm gold}_{R_{\pm}} = 0.164$,
  $z^{\rm gold}_{S} = 0.0718$; $z_{R}^{{\rm silver}} = 0.118$, $z^{\rm silver}_{R_{\pm}} = 0.248$,  $z^{\rm silver}_{S} = 0.020$).}
\label{fig:falpha}
\end{figure}

The approximative $f(\alpha)$-curves calculated from equation (\ref{equ:3z}) 
are compared in Fig. \ref{fig:falpha} with 
those obtained by the usual multifractal formalism \cite{Halsey}.  
Despite of the simplicity of our model there is a good agreement between the
curves. Including linearization around further periodic orbits of the
non-linear map $f^{\pm,0}(E)$ would yield further quantitative improvement. We
consider the $3z$-model as the simplest qualitatively correct approximation. 

Let us note that the  Thouless property $D_{-1} = 1/2$ (i.e. $\tau(-1) = -1$)
yields one linear equation which the three contraction factors have 
to fulfill: 
\begin{equation}
  \frac{z_{R}}{\omega^{d(R)}} + \frac{z_{R_{\pm}}}{\omega^{d(R)}} 
+ \frac{z_{S}}{\omega^{d(S)}} = 1, \qquad \omega = \omega_{\rm gold},
\omega_{\rm silver} 
\end{equation}
Thus, by use of this relation,  
the calculation of the scaling behavior at the band-edges $z_{R}$ and at
the center of the band $z_{S}$ would enable to obtain the missing contraction
factor $z_{R_{\pm}}$. 

\section{Higher quadratic irrational numbers}

Hitherto we have been considering the case of the fixed points of the
Hofstadter rules, where the renormalization always yields the same function
$f^{\pm, 0}_{R(\omega)}(E) = f^{\pm, 0}_{S(\omega)}(E) = f^{\pm,
  0}_{\omega}(E) $  that can be considered as a dynamical system. 
We have shown that  taking into
account not only the fixed points of this dynamical system, but also a
periodic  orbit of length two, satisfactory $f(\alpha)$-curves can be
obtained. 
Now we generalize this approach to the irrational numbers $[0;\overline{n}]$ 
that are not fixed points  of the Hofstadter rules but give rise to a
nontrivial cycle structure that has been shown above. 

In this case we have to deal not only with one function $f^{\pm, 0}_{\omega}(E)$
but with a whole set of function $\{ f^{\pm, 0}_{\omega_{i}}(E) \}$, 
where the $\omega_{i}$ are the irrational numbers occuring in the 
flow diagram of $\omega$. 
In order to calculate the scaling properties of the spectrum, we have to
consider the cycles of the flow diagram. 
To each cycle there is one effective $f^{\pm,0}_{\rm eff}(E)$ that results from
concatenating the corresponding maps $f^{\pm,0}_{\omega_{i}}(E)$ for 
the $\omega_{i}$ occuring in this cycle. For example, the effective map of the
cycle $S^{2}R$ in the case of $\omega = [0;\overline{3}]$ is given by $f_{\rm
  eff}^{\pm} = f_{[0;2,\overline{3}]}^{\pm} \circ
f_{[0;1,\overline{3}]}^{0} \circ f_{[0;\overline{3}]}^{0}$ (consider equation
(\ref{hofstadter}) and the flow diagram). 
Each of these effective maps can be considered as a dynamical system with fixed points 
and periodic orbits of different lengths. 

Now two natural questions arise: Which cycle of the renormalization 
flow does lead to the minimal (respectively maximal) local scaling exponent?  
Which fixed point or periodic orbit of the dynamical system corresponding to 
this cycle does yield the extremal scaling exponent? 
In a first step, these two questions can be addressed numerically by 
determinating the index of the band of maximal (respectively minimal) width
and following its way under the renormalization flow. 

We find that the the minimal scaling exponent for $\omega = [0;\overline{n}]$ 
occurs for the cycle $R$ (these $\omega$ are fixed points of $R$), i.e. the
smallest bands occur at the edges of the spectrum. 
More interestingly, the maximal scaling exponent occurs for the cycle 
$S^{n-1}R$ ($n$ odd) respectively for $S^{\frac{n}{2}-1}R$ ($n$ even). 
The lengths and the structure of these cycles show a striking even-odd effect
which is a direct result of the Hofstadter rules. 
Furthermore, for $n$ odd the maximal scaling exponent is given by the fixed
point of the effective dynamical system, while for even values of $n$ the
asymmetric orbit of length two yields the maximal scaling exponent. 
 
For the minimal exponent  we find for $\omega = [0;\overline{n}]$: 
\begin{equation}
  \alpha_{\rm min} = - \frac{\ln \omegacon}{\ln z_{R}
    ([0;\overline{n}])}, 
\label{amin}
\end{equation}
and for the maximal exponent we obtain 
\begin{eqnarray}
  \alpha_{\rm max} & = & - \frac{2 \ln
    \omegacon}{\ln \left| {f_{\rm eff}^{\pm}}'(E_{1}^{*})\right|}  
\qquad  {\rm if }\quad n \; {\rm odd}\\
  \alpha_{\rm max} & = & - \frac{ \ln
    \omegacon}{\ln \left| {f_{\rm eff}^{\pm}}'(E_{2}^{*})\right|}
\qquad  {\rm if} \quad  n  \; {\rm even},   
\label{amax0}
\end{eqnarray}
where $E^{*}_{1}$ ($E^{*}_{2}$) is the energy of the fixed point (energy of
the orbit of period two) of the effective map $f^{\pm}_{\rm eff}(E)$. 
Replacing the effective map $f^{\pm}_{\rm eff}(E)$ by the individual maps, we 
find 

\begin{eqnarray}
  \alpha_{\rm max} & = & - \frac{2 \ln
    \omegacon}{\sum_{i=1, i\neq 2}^{n} \ln
    z_{S}([0;i,\overline{n}]) + \ln z_{R}([0;2,\overline{n}])}  
\qquad  {\rm if }\quad n \; {\rm odd}\\
  \alpha_{\rm max} & = & - \frac{ \ln
    \omegacon}{\sum_{i=4, i \; {\rm even} }^{n} \ln
    z_{S}([0;i,\overline{n}]) + \ln z_{R_{\pm}}([0;2,\overline{n}])}
\qquad  {\rm if} \quad  n  \;{\rm even}.  
\label{amax}
\end{eqnarray}

\begin{figure}
\centerline{\epsfig{figure=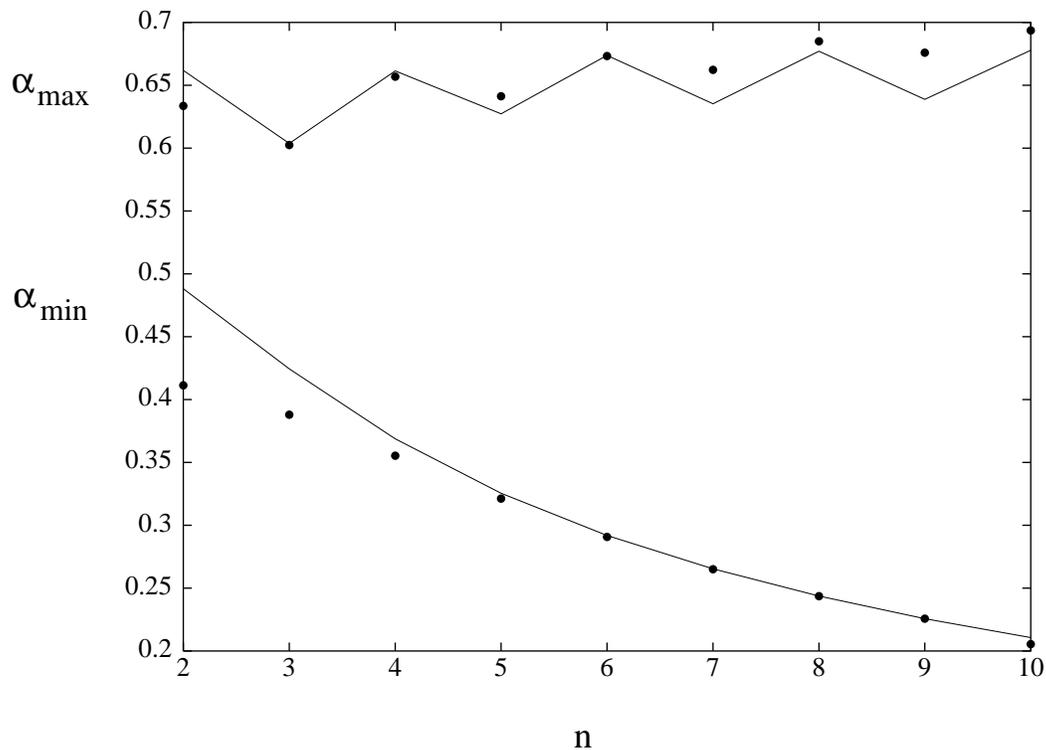,width=10cm,angle=270}}
\caption{The fractal dimensions $D_{-\infty}=\amax$ and $D_{\infty}=
  \amin$ for $\omega = [0;\overline{n}]$ as a function of $n$. 
Results of diagonalization are depicted with points, the lines refer to 
to equation (\ref{equ:amin2}) and (\ref{amax})}
\label{fig:aminmax}
\end{figure}

Inspecting these two equations one is lead to the assumption that $\amax$ 
as a function of $\omega = [0;\overline{n}]$ should be sensible to the parity
of $n$, while $\amin$ is expected to show a monotonic behavior as a function
of $n$.  
This is indeed the case, as shows Fig. \ref{fig:aminmax}. 
For a more quantitative discussion of our approximation, however, the behavior
of $z_{R}$, $z_{R_{\pm}}$ and $z_{S}$ as a function of $\omega$ is
required.  
For limiting cases, as $\omega \to 0$ or $\omega \to \frac{1}{2}$, the
contraction factor $z_{S}$ can be obtained by semiclassical analysis and by
considering tunneling in phase space
\cite{Wilkinson1,Wilkinson3,Rammal,Barelli1,Barelli2,Bellissard},  
if one makes the further 
assumption that $f^{0}_{\omega}(E)$ is nearly linear, so that $z_{S}$ can be
written as ratio of band edges: $z_{S}(\omega) = E_{1-}(\omega)/E_{0}(S(\omega))$. 
For $\omega \to 0$ $z_{S}(\omega)$ is therefore given by the ratio of two
Landau levels and  with the results of \cite{Rammal} we find: 
\begin{equation}
  z_{S}(\omega) = 1 - \pi \omega + \pi \omega^{2} + (2\pi -\pi^{2}
  +\frac{\pi^{3}}{24})\omega^{3} + {\cal O}(\omega^{4}) \quad (\omega \to 0). 
\label{zS1}
\end{equation}
For $\omega \to \frac{1}{2}$ tunneling has to be taken into account for
determining $E_{1-}(\omega)$ and we find
\begin{equation}
-\frac{1}{\ln z_{S}(\omega)} \simeq b_{S}(\frac{1}{2} - \omega) +
c_{S}(\frac{1}{2}-\omega)^{2} +d_{S}(\frac{1}{2}-\omega)^{2} \ln (\frac{1}{2}
- \omega) \quad (\omega \to \frac{1}{2}), 
\label{zS2}
\end{equation}
where $b_{S}$ is given analytically by  $b_{S} =\pi \cdot (\int_{0}^{\pi/2}\ln(\cos k +\sqrt{1+\cos^{2} k }) {\rm d}k)^{-1}
\approx 3.429815 $ and $c_{S}$ and $d_{S}$ are fit parameters (cf. appendix). 
Although equations (\ref{zS1}) and (\ref{zS2}) are only expansions near 
$\omega = 0$ and $\omega = \frac{1}{2}$, they describe $z_{S}(\omega)$ over
the whole range $0 < \omega < \frac{1}{2}$ amazingly well, if one takes
equation (\ref{zS1}) for $0<\omega < 0.25$ and equation (\ref{zS2}) for $0.25
< \omega < 0.5$ (cf. Fig. \ref{fig:zs}). 

In principle, a similar analysis can be performed for the mean slope
$\overline{z_{R}} = (E_{0}(\omega)-E_{1+}(\omega))/(2 E_{0}(R({\omega})))$. 
As the linear approximation for $f^{\pm}(E)$ assumed in this relation is a
rather crude one for general $\omega$, the values of $z_{R}$ and $z_{R_{\pm}}$
can deviate considerably from $\overline{z_{R}}$. This, however, is not the
case for $\omega \to 0$, where $z_{R}$ and $z_{R_{\pm}}$ are well approximated
by $\overline{z_{R}}$ that is given in this regime by 
\begin{equation}
-\frac{1}{\ln \overline{z_{R}}(\omega)} = b_{R}\omega + c_{R}\omega^{2} +
d_{R}\omega^{2} \ln \omega, \quad (\omega \to 0)
\label{zR2}
\end{equation}
where $b_{R} = 2\pi \cdot (\int_{0}^{2\pi} \ln(2-\cos k +\sqrt{(2-\cos
  k)^{2}-1}) {\rm d}k)^{-1}  \approx 0.857454$ and $c_{R}\approx 1.5$ and
$d_{R} \approx 0.37$ again are fit parameters. 
 
As $\amin$ is determined by $z_{R}[0;\overline{n}] \approx \overline{z_{R}}[0;\overline{n}]$ (as $[0;\overline{n}]
\to 0$ for $n \to \infty$),  we insert equation (\ref{zR2}) into (\ref{amin}) 
and obtain  
\begin{equation}
\amin \simeq (b_{R} \omega + c_{R} \omega^{2} + d_{R} \omega^{2} \ln \omega)\ln \omegacon,  
\label{equ:amin2}
\end{equation}
where $\omega = \frac{1}{2}(\sqrt{n^{2}+4} - n)$ and $\omegacon =
\frac{1}{2}(n+\sqrt{n^{2}+4})$. 
Therefore, asymptotically $\amin \sim b_{R} \ln n / n$ ($n \to \infty$)
and $\lim_{n \to \infty} \amin = 0$. 

For $\amax$ the values $z_{R}$ and $z_{R_{\pm}}$ for $\omega$ near to
$\frac{1}{2}$ are needed. As they are not directly available in our approach,
we take approximately $z_{R}(\omega) = \overline{z_{R}}(\omega)$ and
$z_{R_{\pm}} = 1.4\overline{z_{R}}(\omega)$, where $\overline{z_{R}}(\omega)$
is given by equation (\ref{zR2}). This approximation is intended to give the
qualitative behavior for $\amax$. 

In Fig. \ref{fig:aminmax} the predictions for $\amin$ and $\amax$ for $3\le 
n\le 10$ are compared with the corresponding values obtained by 
diagonalization. 
For $\amin$ and $n \ge 5$  there is  good agreement between pediction and
simulation, while for $n <5$ the predicted value of $\amin$ are to high,
reflecting the fact that $z_{R} < \overline{z_{R}}$. 
For $\amax$ we find qualitatively the even-odd oscillations, but of course,
due to the problems mentionned above, we can not hope to achieve quantitative
agreement. In particular an open question is the behavior of $\amax$ for $n
\to  \infty$. For the moment, we can only state, that if $\ln z_{R}(\omega)$
and $\ln z_{R_{\pm}}(\omega)$ is bounded for $\omega \to \frac{1}{2}$, then
$\amax$ does not reach the value 1 for $n \to \infty$.   \\

Summing up, we have used a refined version of Hofstadter rules together 
with results from semiclassical analysis to link the
maximal and minimal fractal dimension of the spectrum of the Harper equation
to the continued-fraction expansion of the irrational number characterizing
the flux per unit-cell. 
We have shown that the oscillatory behavior of $\amax$ for
$\omega=[0;\overline{n}]$ as a function of $n$ can be 
qualitatively explained by the discrete flow diagrams of the Hofstadter rules.
For $\amin$ an explicit expression has been given which describes fairly well 
the data obtained by diagonalization. 
We conjecture that the result for $\amax$ can be generalized to 
quadratic numbers of the
continued fraction expansion $\omega =
[0;\overline{n_{1},n_{2},\cdots,n_{p}}]$. Take for example  
$\omega=[0;\overline{n_{1}n_{2}}]$. In this case the renormlization flow
diagramm contains $\frac{1}{2}(n_{1}+n_{2})$ different irrational numbers, if
both $n_{1}$ and $n_{2}$ are even. In all the other cases, there are
$n_{1}+n_{2}$ different irrational numbers. 
Thus for even values of $n_{1}$ the cycle structure depends on the parity of
$n_{2}$, while this is not the case for $n_{1}$ odd. Hence, for fixed even 
values of $n_{1}$ we expect even-odd oscillations of $\amax$ as function of
$n_{2}$, while for odd values of $n_{1}$ no such oscillations are
expected.  

Finally we note that the applicability of the Hofstadter rules to a wide class
of quasiperiodic models suggests that our method could be useful for other 
models than the one we have considered here.

\section{Appendix}

This appendix is intended to provide the calculations for $z_{S}$ and
$\overline{z_{R}}$ in the semiclassic approximation using 
methods and results described in 
\cite{Wilkinson3,Rammal,Barelli1,Barelli2,Bellissard}.  

Equation (\ref{zS1}) can be obtained directly by inserting the semiclassical
expansion of the Landau levels up to order three \cite{Rammal} into 
\begin{equation}
z_{S}(\omega) = \frac{E_{1-}(\omega)}{E_{0}(\frac{\omega}{1-2\omega})}, \quad
(\omega \to 0).   
\end{equation}
For $\omega \to \frac{1}{2}$ the contraction factor $z_{S}$ is given by 
\begin{equation}
z_{S}(\omega) = \frac{E_{1-}(\omega)}{E_{0}(4(\frac{1}{2}-\omega))}, \quad
(\omega \to \frac{1}{2}).     
\label{app:zS}
\end{equation}
While the denominator corresponds to Landau levels near $\omega = 0$ and the
already mentioned semiclassical expansion can be used, the numerator involves
the calculation of $E_{1-}(\omega)$ for $\omega \to \frac{1}{2}$.  
Since for $\omega = \frac{1}{2}$ the effective Hamiltonian 
$H = \pm 2 \sqrt{\cos^{2}k_{1} +  \cos^{2}k_{2}}$ has the form of a Dirac 
operator at the four critical points $(\pm \frac{\pi}{2},\pm \frac{\pi}{2})$
(independent signs) 
\cite{Rammal}, we are left with the calculation of tunneling between Dirac 
levels. 

\begin{figure}
\centerline{\epsfig{figure=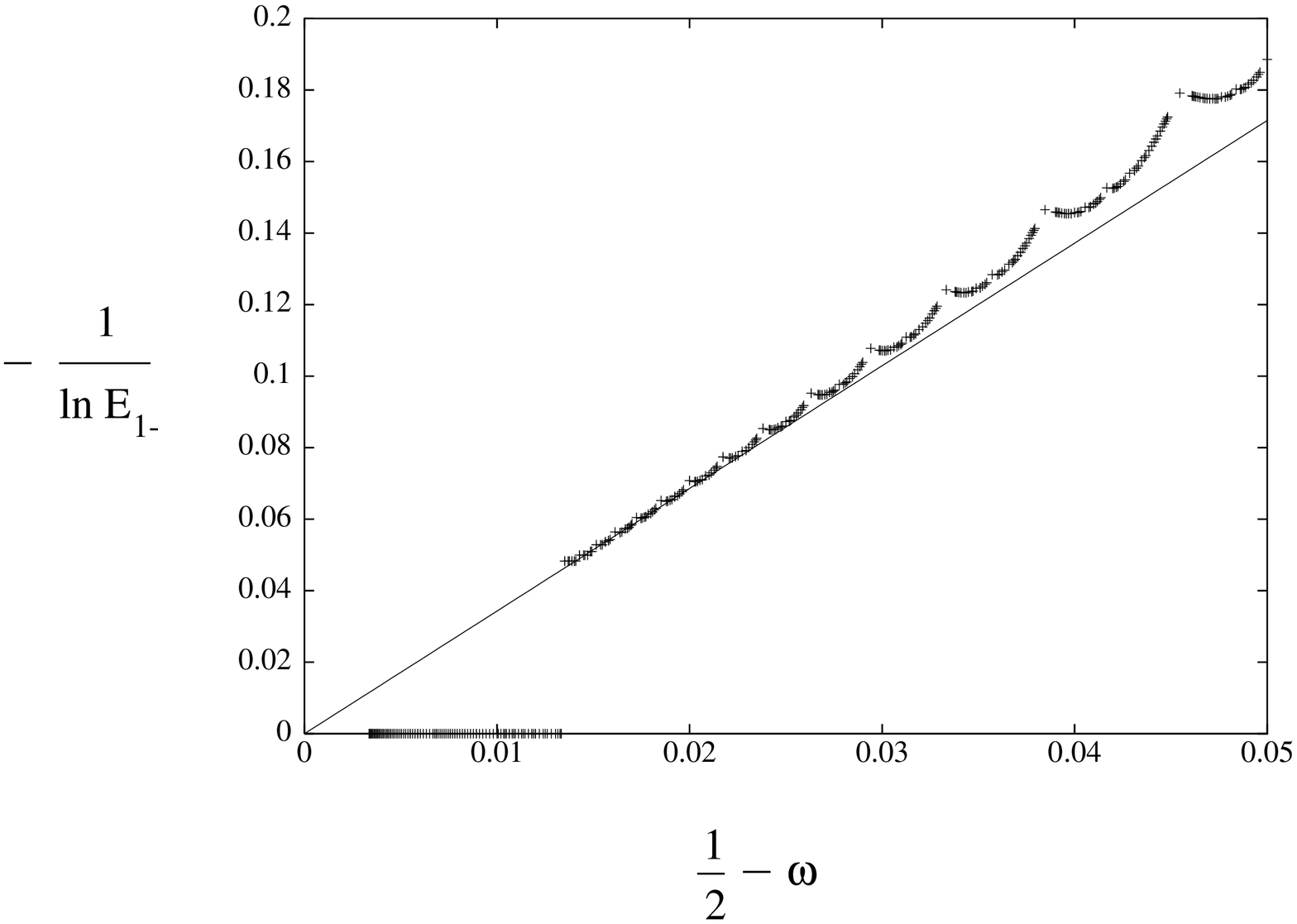,width=10cm,angle=270}}
\caption{The broadening of the Dirac level at $E=0$, $\omega =\frac{1}{2}$
  compared to the result from tunneling in phase space $-1/\ln E_{1-}(\omega)
  \sim 2 \pi (\frac{1}{2}-\omega)/|{\rm Im }S|$, where the imaginary part of
  the 
  action is given by $|{\rm Im}S |= \int_{-\pi/2}^{\pi/2}\ln(\cos k
  +\sqrt{1+\cos^{2} k}){\rm d}k \approx 1.831931$. For $1/2 -\omega < 0.013$
  the broadening of Dirac levels can not be resolved numerically.}
\label{fig:dirac}
\end{figure}

Although the usual formula for level splitting due to tunneling
\begin{equation}
  \Delta E \sim {\rm const} \cdot  \hbar  \cdot \exp\left(-\frac{|{\rm Im} S|}{\hbar}\right), 
\label{action1}
\end{equation}
where $S$ is the action integral between degenerate minima, $\hbar$ is
Planck's constant and the proportional constant is related to the classical
frequency, 
was originally derived for tunneling between parabolic minima \cite{Landau},
we tentatively use it for our case, where the minima are conical.  
With $\hbar = 2\pi (\frac{1}{2}- \omega)$ \cite{Rammal} we obtain 
\begin{equation}
  E_{1-}(\omega) \sim (1/2- \omega)\exp \left(- \frac{|{\rm Im} S|}{2\pi
    (1/2- \omega)}\right), 
\label{E1m}
\end{equation}
where the action integral has to be calculated between two adjacent minima in
phase space. 
The (complex) orbits are given by $H=0$, i.e. $\cos^{2} k_{1} + \cos^{2}
k_{2} = 0$ and therefore 
\begin{eqnarray}
  S & = &  \int_{-\pi/2}^{\pi/2} \arccos (\pm {\rm i} \cos k_{1}){\rm d} k_{1} \\
 |{\rm Im} S|   & = &  2 \int_{0}^{\pi/2} \ln(\cos k_{1} +
    \sqrt{1+\cos^{2}k_{1}}){\rm d} k_{1} \approx 1.831931.   
\label{action2}
\end{eqnarray}
Formula (\ref{E1m}) with (\ref{action2}) inserted is compared to numerical
results from diagonalisation in Fig. \ref{fig:dirac}.  We note that the fact 
of the action being purely imaginary shows the absense of braiding of the
Dirac levels.  
To calculate $z_{S}(\omega \to 1/2)$ we insert (\ref{E1m}) together with 
\begin{equation}
  E_{0}(4(1/2-\omega)) \sim 4 - 8 \pi (1/2 - \omega)  +{\cal O}\left((1/2-\omega)^{2}\right)
\end{equation}
 into (\ref{app:zS})
and obtain (\ref{zS2}). The parameters $c_{S} \approx 9.6$ and $d_{S}\approx
5.2$  are determined by fitting to the numerical data. 

The calculations for $\overline{z_{R}}(\omega \to 0)$ are analogous, but only
Landau levels have to be considered.

\newpage 

\begin{figure}
\begin{equation}
  \begin{array}{|l|l|c|} \hline  
\mbox{\rule[-2mm]{0cm}{0.6cm}} & \mbox{elementary cycles $C$} & \mbox{generation  loss $d(C)$} \\ \hline \hline 
\mbox{\rule[-2mm]{0cm}{0.8cm}} n \textrm{ even} & S^{j}R,\quad 0 \le j \le \frac{1}{2}n-1 & 1 \\
                 & S^{n/2}  & 2\\ \hline
\mbox{\rule[-2mm]{0cm}{0.8cm}} n \textrm{ odd} & S^{j}R, \quad 0 \le j \le \frac{1}{2}(n-3) & 1\\
                & S^{j}R, \quad \frac{1}{2}(n-1) \le j \le n-1 & 2\\
& S^{n} & 3 \\ \hline
  \end{array}
\end{equation}
\caption{\mbox{Cycles occuring in the discrete renormalization diagram for $\omega =
  [0;\overline{n}]$.}\hspace*{10cm}}
\label{tableau}
\end{figure}

\newpage

\begin{figure}
\centerline{\epsfig{figure=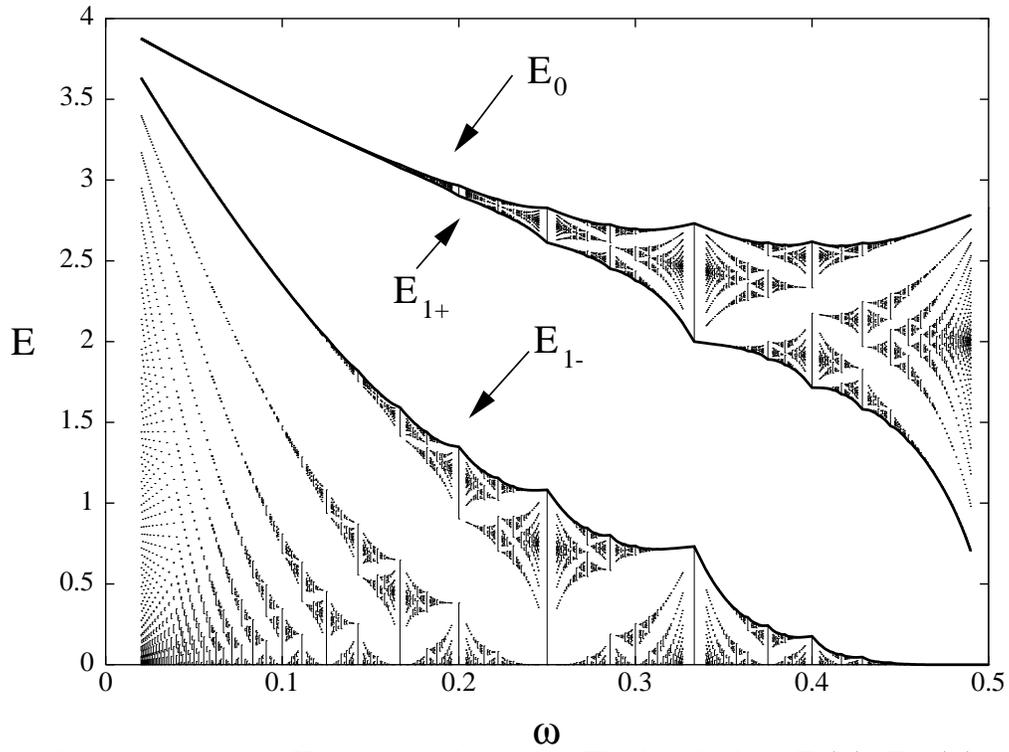,width=10cm,angle=270}}
\caption{One quarter of the Hofstadter's butterfly. The band edges
  $E_{0}(\omega)$,  $E_{1+}(\omega)$ and $E_{1-}(\omega)$ are depicted with
lines.}  
\label{fig:edges}
\end{figure}  
\newpage

\newpage

\begin{figure}
\centerline{\epsfig{figure=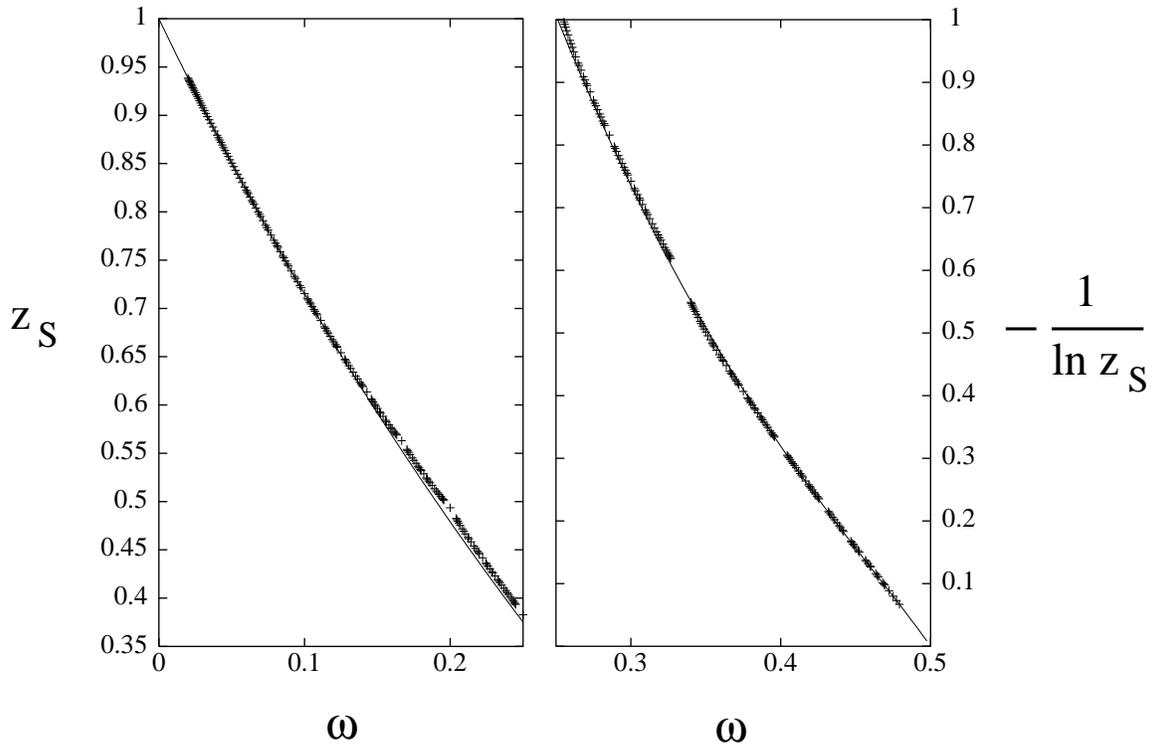,width=10cm,angle=270}}
\caption{The contraction factor $z_{S}(\omega)=E_{1-}(\omega)/E_{0}(\omega)$
  compared to the results from semiclassical analysis (equation (\ref{zS1})
  for $0 < \omega < 0.25$ and equation (\ref{zS2}) for $0.25 < \omega < 0.5$
  ($b_{S}= 3.429815$, $c_{S} = 9.6$, $d_{S} = 5.2$)). Note that $z_{S}(\omega)$
  is remarkably smooth compared to $E_{1-}(\omega)$ and $E_{0}(\omega)$
  (Fig. \ref{fig:edges}).} 
\label{fig:zs}
\end{figure}  
\newpage

\newpage

\begin{references}
\bibitem{Harper} Harper P G 1955 
{\it Proc. Roy. Soc. London A} {\bf 68}, 874 
\bibitem{Thouless0} Thouless D J, Kohmoto M, Nightingale P and den Nijs M 
1982 
{\it Phys. Rev. Lett.} {\bf 49}, 405 
\bibitem{Pannetier} Pannetier B, Chaussy J, Rammal R and  Villegier J C 1984 
{\it Phys. Rev. Lett.} {\bf 53}, 1845  
\bibitem{Gerhardts} Gerhardts R , Weiss D and Wulf U 1991 
{\it Phys. Rev. B} {\bf 43}, 5192 
\bibitem{Rammal} Rammal R and Bellissard J 1990 
{\it J. Phys. France} {\bf 51}, 1803 
\bibitem{Bellissard0} Bellissard J 1987 in 
{\it Operator Algebras and  applications} Vol.2,  editors: Evans D E and Takesaki M, Cambridge University Press 
\bibitem{Wiegmann} Wiegmann P B and Zabrodin A V 1994  
{\it Phys. Rev. Lett.} {\bf 72}, 1890
\bibitem{Schlosser} Schl\"osser T, Ensslin K, Kotthaus J P and Holland M 1996 
{\it Europhys. Lett.} {\bf 33}, 683 
\bibitem{Halsey} Halsey T S, Jensen M H, Kadanoff L P and  Shraiman B I 1986 
{\it Phys. Rev. A} {\bf 33}, 1141 
\bibitem{Kohmoto1} Kohmoto M and Oono Y 1984 
{\it Phys. Lett. A} {\bf 102}, 145 
\bibitem{Tang} Tang C and Kohmoto M 1986 
{\it Phys. Rev. B} {\bf 34}, 2041  
\bibitem{Kohmoto2} Kohmoto M, Sutherland B and  Tang C 1987 
{\it Phys. Rev. B} {\bf 35}, 1020  
\bibitem{Zheng} Zheng W M 1987 
{\it Phys. Rev. A} {\bf 35}, 1467 
\bibitem{Hiramoto} Hiramoto H and Kohmoto M 1992 
{\it Int. J. Mod. Phys.} {\bf 6}, 281
\bibitem{Ikezawa} Ikezawa K and Kohmoto M 1994
{\it J. Phys. Soc. Japan} {\bf 63}, 2261 
\bibitem{Fred1} Pi\'echon F, Benakli M and Jagannathan A 1995  
{\it Phys. Rev. Lett.} {\bf 74}, 5248 
\bibitem{Rudinger} R\"udinger A and Sire C 1996 
{\it J. Phys. A} {\bf 29}, 3537
\bibitem{Bell} Bell S C  and  Stinchombe R B 1989 
{\it J. Phys. A} {\bf 22}, 717 
\bibitem{Thouless} Thouless D J 1983 
{\it Phys. Rev. B} {\bf 28}, 4272 
\bibitem{Last} Last Y and Wilkinson M 1992 
{\it J. Phys. A} {\bf 25}, 6123
\bibitem{Tan} Tan Y 1995 
{\it J. Phys. A} {\bf 28}, 4163
\bibitem{remark} 
Using however the thermodynamic formalism for the multifractal properties
 $\sum_{i=1}^{N} \Delta_{j}^{(1-q)D_{q}} \sim N^{q}$ \cite{Halsey},  
the Thouless property\cite{Thouless,Tan,Last} for the scaling of the total 
bandwith $\sum_{i=1}^{N}\Delta_{j} \sim N^{-1}$ gives immediately 
$D_{-1} = 1/2$, and therefore $D_{0}<D_{-1}=1/2$, provided $D_{q}$ is strictly
monotonically decreasing, that is the generic case for a multifractal.  
\bibitem{Wilkinson2} Wilkinson M  and Austin E J 1994 
{\it Phys. Rev. B} {\bf 50}, 1420   
\bibitem{Fred2} Pi\'echon F 1996 
{\it Phys. Rev. Lett.} {\bf 76}, 4372 
\bibitem{Holzer1} Holzer M 1988 
{\it Phys. Rev. B} {\bf 38}, 1709 
\bibitem{Holzer2} Holzer M 1988 
{\it Phys. Rev. B} {\bf 38}, 5756 
\bibitem{Hofstadter} Hofstadter D R 1976 
{\it Phys. Rev. B} {\bf 14}, 2239 
\bibitem{thesepiechon} Pi\'echon F 1995, PhD Thesis (Paris)
\bibitem{Wilkinson1} Wilkinson M 1987 
{\it J. Phys. A} {\bf 20}, 4337 
\bibitem{Wilkinson3} Wilkinson M and Austin E J 1990 
{\it J. Phys. A} {\bf 23}, 2529
\bibitem{Barelli1} Barelli A and Kreft C 1991 
{\it J. Phys. I France} {\bf 1}, 1229 
\bibitem{Barelli2} Barelli A and Fleckinger R 1992 
{\it Phys. Rev. B} {\bf 46}, 11559 
\bibitem{Bellissard} Bellissard J and Barelli A 1993 
{\it J. Phys. I France} {\bf 3}, 471 
\bibitem{Landau} Landau L D and Lifshitz E M 1955 
{\it Quantum Mechanics}, Pergamon (Oxford)
\end{references}
\end{document}